\pdfoutput=1
\documentclass[11pt]{article}

\usepackage[preprint]{acl}
\usepackage{fancyvrb}

\usepackage[utf8]{inputenc} % allow utf-8 input
\usepackage[T1]{fontenc}    % use 8-bit T1 fonts
\usepackage{hyperref}       % hyperlinks
\usepackage{url}            % simple URL typesetting
\usepackage{booktabs}       % professional-quality tables
\usepackage{amsfonts}       % blackboard math symbols
\usepackage{nicefrac}       % compact symbols for 1/2, etc.
\usepackage{microtype}      % microtypography
\usepackage{amsmath}

\usepackage{longtable}

\usepackage{multirow}
\usepackage{subcaption}

\usepackage{comment}
\usepackage{tikz}
\usepackage{pifont}
\usepackage{tablefootnote}
\usepackage{wrapfig}
\usepackage{paralist}
\usepackage{ulem}
\usepackage{siunitx}
\usepackage{arydshln}
\usepackage{makecell}
\usepackage{xcolor,colortbl}
\usepackage{tcolorbox}

\usepackage{xspace}

\usepackage{listings}
\usepackage{adjustbox}

\usepackage{enumitem}

\usepackage{amssymb}% http://ctan.org/pkg/amssymb
\usepackage{pifont}% http://ctan.org/pkg/pifont
\newcommand{\xmark}{\ding{55}}%
\definecolor{MyColor}{RGB}{50, 100, 250}
\definecolor{Orange}{RGB}{244, 101, 66}
\definecolor{Red}{RGB}{255, 0, 0}
\definecolor{Green}{RGB}{0, 255, 0}
\definecolor{Blue}{RGB}{0, 0, 255}
\definecolor{codegreen}{rgb}{0,0.6,0}
\definecolor{codegray}{rgb}{0.5,0.5,0.5}
\definecolor{codepurple}{rgb}{0.58,0,0.82}

\newcommand{\tool}{\textsc{ZPS}\xspace}

\title{Solving Zebra Puzzles \\ Using Constraint-Guided Multi-Agent Systems}

\author{Shmuel Berman \\
  \texttt{sb6870@princeton.edu} \\\And
  Baishakhi Ray \\
  \texttt{rayb@cs.columbia.edu} \\\And
    Kathleen McKeown \\
  \texttt{kathy@cs.columbia.edu} \\}

\begin{document}
\maketitle
\begin{abstract}
Prior research has enhanced the ability of Large Language Models (LLMs) to solve logic puzzles using techniques such as chain-of-thought prompting or introducing a symbolic representation. These frameworks are still usually insufficient to solve complicated logical problems, such as Zebra puzzles, due to the inherent complexity of translating natural language clues into logical statements.
%We introduce a multi-agent system that integrates LLMs with an off the shelf theorem prover that addresses the difficulty of this task by partitioning the problem-solving process into discrete, manageable components and utilizing feedback between agents to allow them to refine their own answers repeatedly. 
We introduce a multi-agent system, the Zebra Puzzle Solver (\tool), that integrates LLMs with an off the shelf theorem prover. This system tackles the complex puzzle-solving task by breaking down the problem into smaller, manageable parts, generating 
%\kmnote{I think we really need to define SMT at some point in time. Could we put it in parens above after "off the shelf theorem prover"? That would help for me and I suspect other NLP readers.}
SMT (Satisfiability Modulo Theories) code  to solve them with a theorem prover, and using feedback between the agents to repeatedly improve their answers.
We also introduce an automated grid puzzle grader to assess the correctness of our puzzle solutions and show that the automated grader is reliable by evaluating it in a user-study. Our approach shows improvement in all three LLMs we tested, with GPT-4 showing 166\% improvement in the number of fully correct solutions.
\end{abstract}

\section{Introduction}

Automated problem solving has long been a major goal in the field of Artificial Intelligence. This task ranges from trivial problems, like simple arithmetic or string searches, to more complex ones, such as solving a chess position. However, unstructured problems presented in natural language introduce additional complications in modeling the problem accurately. Solving such problems has been extensively studied, from simple mathematical problems in the subfield of word problem solving to applications like automated code generation by Large Language Models (LLMs)~\citep{mukherjee2008,chen2021}. These problems are particularly difficult because translating natural language into a precise logical or computational form requires sophisticated understanding and interpretation, making it a significant challenge in AI research.

\begin{figure}[h]
    \centering
    \includegraphics[width=0.45\textwidth]{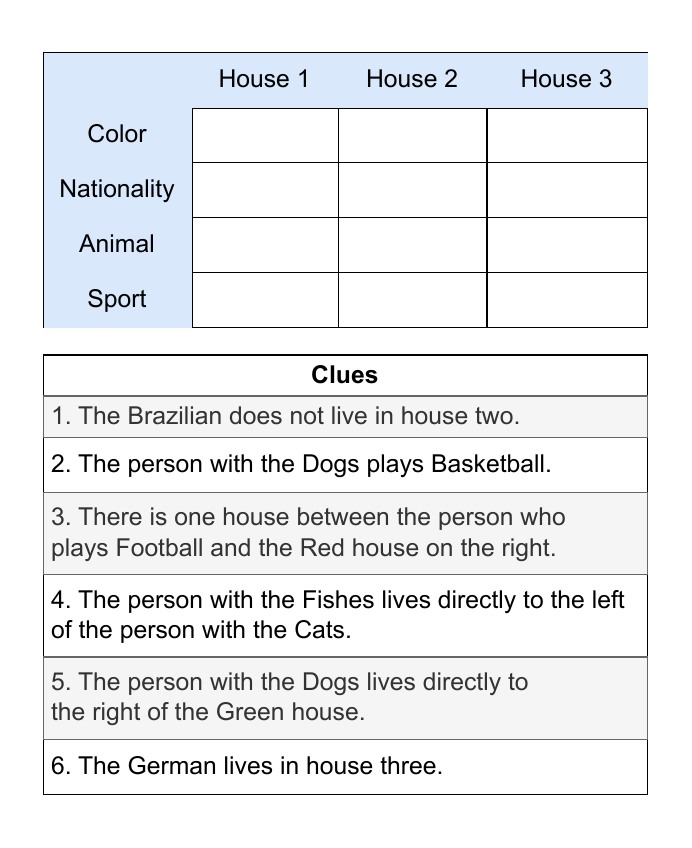}
    \caption{An Example Zebra Puzzle.}
    \label{fig:example}
\end{figure}

In this paper, we focus on a particular type of unstructured natural language problem known as a logic grid problem, or colloquially, an Einstein or Zebra puzzle. 
A Zebra puzzle is a set of natural language assertions involving multiple entities that are linked by various attributes (Fig.~\ref{fig:example} shows an example).
To solve a puzzle, the user must correctly assign attributes to all of the entities. These attributes range from descriptions to relative ordering. Participants are provided with a series of clues in natural language, which they must use to deduce the correct relationships using logical reasoning and by adhering to implicit domain constraints. These puzzles require the solver to map from natural language to structured space, understand implicit 
%KM - changed "need" to "use" to keep parallel structure in teh conjunction
assumptions, and in some cases 
%need
use domain-specific knowledge. 
For instance, as illustrated in Fig.~\ref{fig:example}, the solver must assign the correct attributes for three houses based on a series of interconnected clues.

Zebra puzzles are particularly challenging due to:
\begin{itemize}%[leftmargin=*,topsep=0pt,itemsep=-1ex]
    \item Complex Inferences: Each clue provides partial information that must be combined with others to deduce the solution.
    \item High Interdependency: 
   % Misinterpreting  
    An error on one clue 
   % can 
    significantly impacts others, making the solution space highly interconnected.
    \item Natural Language (NL) Clues: Translating ambiguous NL clues
 %   , often ambiguous, 
    into logical statements or formal representations is challenging.
    \item Large Solution Space: The solver needs to explore numerous possibilities and combinations 
    %KM - this doesn't fit here and causes a line with one word. It could be moved into bullet 1 or the last bullet. I put it in the last bullet as it can there without adding space.
 %   (can be domain specific)  
    to find the solution.
    \item Consistency Checking:
    %KM cut to avoid widow
    %Each 
    Potential solutions must be checked against all clues, 
    which is computationally intensive and requires sophisticated, domain-specific reasoning.
\end{itemize}
The factors mentioned make Zebra puzzles difficult for both humans and AI systems due to the need for precise interpretation, inference, and logical reasoning. In Fig.~\ref{fig:example}, for example, a solver cannot simply map the spatial relationship between the Football house and the Red house. It must also encode additional constraints: the Football and Red houses occupy House 1 or House 3, respectively, and they must not be the same house. Encoding these constraints is non-trivial, as it requires detailed semantic interpretation of the clue's subtext. Failure to accurately encode these subtleties usually renders the puzzle unsolvable. 

This complexity has empirically been shown to challenge puzzle-solving models significantly. 
%\kmnote{Need a citation for this sentence if we include it. Or, move this sentence to the beginning of the next paragraph and then the citations are clear. I've done this for now. }
Prior work often employed human-in-the-loop methods. \citet{milicevic2012puzzler} translated puzzles into formal logic but required users to rephrase or rewrite ambiguous clues. \citet{claes2019user} developed ZebraTutor, which creates a puzzle-specific lexicon to formalize the problem but needed users to edit the lexicon for accuracy. Prior research using ChatGPT to solve Zebra puzzles reported a correctness rate of only 8.33\% \citep{groza2023}, with performance deteriorating significantly as the problem's complexity increases.

%KM Rephrased for fluency
%Due to the complex nature of Zebra puzzles, 
Due to their complexity, solving Zebra puzzles  effectively requires the use of a constraint solver; 
%as 
a solver can efficiently determine
%figure out efficiently 
%what is 
the feasible and infeasible solution space within the given constraints. However, converting natural language clues into a formal representation suitable for a solver is a non-trivial task. This process often involves intricate 
%KM Cut this part to remove the widow at end of paragraph. Interpretation alone conveys the meaning.
%translation and 
interpretation of clues, which must be precise to ensure that the solver can operate correctly. Additionally, maintaining consistency across all clues requires iterative back-and-forth reasoning.

To address the challenges inherent in solving Zebra puzzles, we introduce a multi-agent based system, \tool. This system decomposes the problem-solving process into discrete, manageable components, enhancing the handling of complex interdependencies and constraints. Each agent is responsible for a specific aspect of the problem, working collaboratively and using feedback loops to refine their answers and ensure consistency.

In this framework, we conceptualize integrating Large Language Models (LLMs) with formal reasoning. 
First, an LLM agent decomposes a given puzzle into sub-problems. Then, another LLM agent interprets NL clues of each sub-problem and generates SMT-LIB translations of the constraints and parameters. 
%\kmnote{What's missing in this description is the process of decomposing/planning into smaller subtasks. I think it should be stated here somewhere as it is part of the contributions. Perhaps this is why there is more than 1 LLM?}
An off-the-shelf SMT solver~\footnote{Satisfiability Modulo Theories (SMT) is a decision problem that involves determining whether a given logical formula is satisfiable, considering various background theories like Arithmetic, Arrays, Bit-Vector, etc. SMT extends the concept of Boolean satisfiability (SAT) by incorporating more complex theories.} then processes these translations to produce a model that corresponds to the solution. 
The output, including the model and any syntactic errors, is fed back to the LLM which generates a new translation addressing syntactic and semantic errors, emulating back and forth reasoning.
This continuous feedback refines the model’s predictions and ensures the translations are both syntactically correct and solvable. 
To this end, our approach demonstrates improvements across all three LLMs we tested, with GPT-4 showing up to a 166\% increase in the number of fully correct solutions.
The main contributions of our research are as follows:

% \begin{enumerate}[leftmargin=*,topsep=0pt,itemsep=-1ex]
%     \item We demonstrate that our LLM agent-based approach for solving Zebra Puzzles phrased in English significantly improves upon existing baseline methodologies.
%     \item We implement a plan generation and decomposition strategy, enabling step-by-step reasoning that enhances the solving process, especially for weaker LLMs.
%     \item We introduce an iterative conversation-based feedback mechanism that allows for continual refinement of solutions, adapting dynamically to the solving context.
%     \item We incorporate an autograder within our system to evaluate the accuracy of solutions, ensuring reliability and precision in automated assessments. We also present the results of a user study showing that this autograder correlates very well with human graders.
%     \item We extend the utility of LLMs by applying these techniques to generate SMT-LIB code, providing a formal environment for the LLM to operate in.
% \end{enumerate}
%\rayb{Summarize the result} \kmnote{I like these contributions. }
\begin{enumerate}%[leftmargin=*,topsep=0pt,itemsep=-1ex]
    \item We demonstrate that combining a formal constraint solver with an LLM interpreter using an agent-based approach for solving Zebra Puzzles significantly improves upon existing baseline methodologies.
    \item We implement a plan generation and decomposition strategy, enabling step-by-step reasoning that enhances the solving process.
    \item We introduce an iterative conversation-based feedback mechanism that allows for continual refinement of solutions, adapting dynamically to the solving context.
    \item We incorporate an autograder within our system to evaluate the accuracy of solutions, ensuring reliability and precision in automated assessments. We also present the results of a user study showing that this autograder correlates very well with human graders.
    %\item We extend the utility of LLMs by applying these techniques to generate SMT-LIB code, providing a formal environment for the LLM to operate in.
\end{enumerate}

Our code is available \href{https://github.com/shmublu/anon_emnlp/tree/main}{here}.

\section{Methodology}

\begin{figure}[t!]
    \centering
    \includegraphics[width=0.5\textwidth]{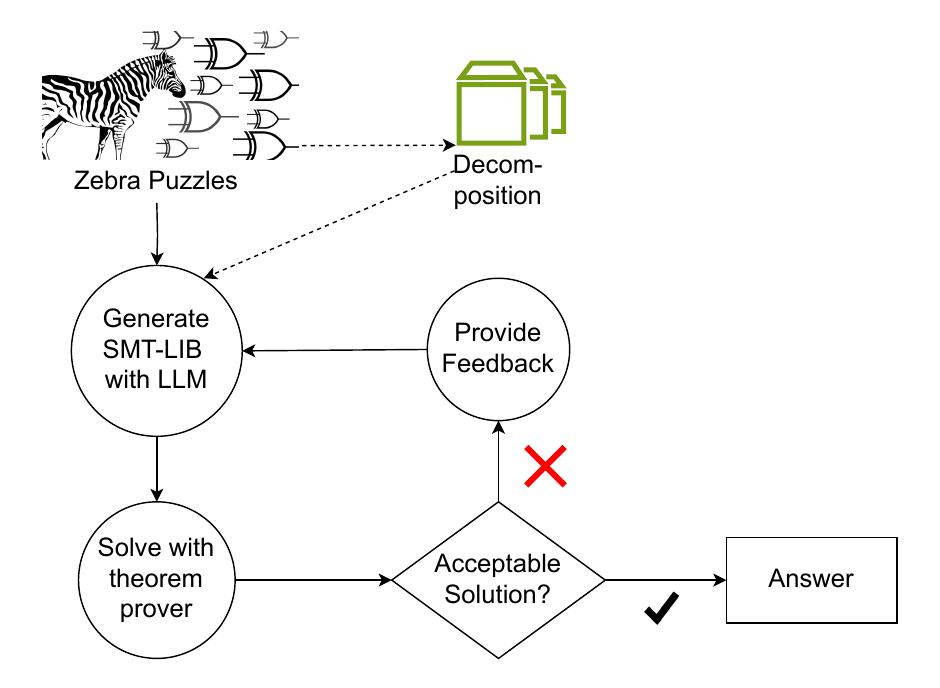}
    \caption{Logic Puzzle Solver Workflow}
    \label{fig:logicpuzzle}
\end{figure}

\begin{figure*}[t!]
    \centering
    \includegraphics[width=\textwidth, height=6cm]{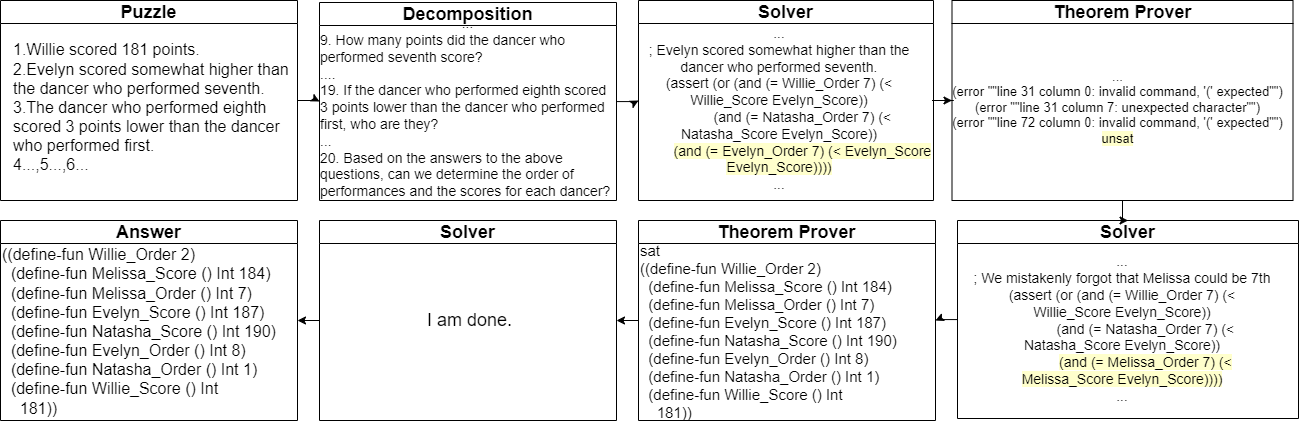}
    \caption{Example Feedback Puzzle Solving Process. The puzzle is decomposed and then the LLM-agent attempts to translate it into a logical SMT formula. The theorem prover attempts to solve it, and the feedback is fed back into the LLM-agent so that it can modify its formal representation.}
    \label{fig:logicpuzzlefeedback}
\end{figure*}

%\subsection{Overview}

We integrate LLMs with formal systems within a multi-agent framework to solve Zebra puzzles. 
The process involves a series of steps where the problem is decomposed, translated into a formal language (SMT-LIB), solved using a theorem prover, and iteratively refined based on feedback. This approach aims to leverage the strengths of both LLMs and formal solvers, ensuring robust problem-solving capabilities.

\subsection{Multi-Agent Workflow}

The workflow, as illustrated in Figure~\ref{fig:logicpuzzle},  integrates multiple agents to transform a natural language puzzle into a logically solvable structure and then iteratively refines the solution. The process is initiated by the Decomposition Agent and continuously refined through a feedback loop that encompasses both the translation to SMT-LIB and the solving phases. Figure~\ref{fig:logicpuzzlefeedback}
shows a working example of puzzle solving by our method. 

\paragraph{Decomposition} The input puzzle, expressed in natural language, is first decomposed by the Decomposition LLM-Agent. This agent identifies and isolates key entities, attributes, and relationships, structuring them into smaller, systematically translatable components. This is a first step that ensures the puzzle is presented in a format amenable to formal processing.
    
\paragraph{Feedback Loop}
    The core of our methodology lies in the feedback loop where continuous refinement of the solution occurs. This loop integrates the translation of decomposed components into SMT-LIB format by the Solver LLM-Agent and the subsequent problem solving using a theorem prover, whose output serves as feedback. Each iteration through the loop consists of the following steps:
    \begin{itemize}[leftmargin=*,topsep=0pt,itemsep=-1ex]
        \item \textbf{Translation to SMT-LIB}: After decomposition, the puzzle components are systematically translated into SMT-LIB (Satisfiability Modulo Theories Library) by the Solver LLM-Agent. This format is essential for interfacing with theorem provers and ensures that logical constraints and relationships are accurately represented.
        \item \textbf{Solving with Theorem Prover}: The SMT-LIB formatted components are then processed by the theorem prover (Z3 in our implementation). The theorem prover attempts to find a satisfying assignment that adheres to all given constraints.
        \item \textbf{Evaluate and Refine}: The solution generated by the theorem prover is evaluated by the Solver LLM-Agent to determine if it meets the puzzle's requirements. If the solution is deemed insufficient-- either due to explicit errors or because of how the attributes are assigned-- modifications are made to the translation of the SMT-LIB formalization of the puzzle and the cycle repeats. Otherwise, the LLM-agent submits its final answer.
    \end{itemize}

This iterative process ensures that the Solver LLM-Agent and the Theorem Prover continually refine the solution until the Solver LLM-Agent is satisfied with the final assignments.

% \subsubsection{Agents}
% We define the following agents that interact in the puzzle solving workflow(see Fig \ref{fig:logicpuzzle}):

% \paragraph{Solving Agent} The primary agent responsible for solving puzzles using the feedback loop with z3.

% \paragraph{Decomposition Agent} An optional agent that breaks down the original puzzle into simpler clues, making them easier for the solving agent to process.

% \paragraph{Consistency Checker Agent} Evaluates if the final solution aligns with the original clues. Subsumed into the role of the solving agent.
%In some experimental configurations, the decomposition and consistency agents are explicitly defined. If not explicitly defined, these responsibilities are assumed to be implicitly handled by the primary solving-agent's workflow.

%\subsection{Modeling the Agent Environment}
\subsection{Modeling the Agent Environment}

The feedback loop is how the agents engage with each other. This loop is mathematically modeled using a combination of evaluation functions and error detection mechanisms, which together guide the system towards a solution that optimally satisfies the problem constraints.

%\textbf{Formal Definition:} 
More formally, let \( \mathcal{D}, \mathcal{G}, \mathcal{T}, \mathcal{E} \), and \( \mathcal{F} \) represent the decomposition, translation to SMT-LIB, theorem solving, evaluation, and feedback functions, respectively. The feedback loop can be described by the following recursive function:
%\kmnote{This feels repetitive though more forma. It takes up a lot of space. Do we need it/ }

\[
S_{k+1} = \mathcal{F}(\mathcal{E}(\mathcal{T}(\mathcal{G}(\mathcal{D}(P)), S_k)), S_k)
\]

Where,\\ 
%\kmnote{I was so traumatized by having a previoius paper desk rejected by playing with space in the itemize command that I would not do it even here. I think we could get around by not using itemize (so no bullets will appear and each variable is on a separate line. I've make an attempt. see what you think. }\\
%\begin{itemize}[topsep=0pt,itemsep=-1ex]
%    \item \( P \): initial puzzle in natural language.
%    \item \( S_k \):  solution state at the \( k \)-th iteration.
%    \item \( \mathcal{D}(P) \): decomposes \( P \) into a structured format amenable to translation.
%    \item \( \mathcal{G} \): translates this structure into the SMT-LIB format.
%    \item \( \mathcal{T} \): applies the theorem prover to find a solution that satisfies the logical constraints.
 %   \item \( \mathcal{E} \): evaluates this solution to determine its adequacy in solving the puzzle's clues.
 %   \item \( \mathcal{F} \): adjusts the translation based on the evaluation, aiming to correct any errors or optimize the solution.
%\end{itemize}
%\begin{itemize}[topsep=0pt,itemsep=-1ex
\( P \): initial puzzle in natural language.\\
\( S_k \):  solution state at the \( k \)-th iteration.\\
 \( \mathcal{D}(P) \): decomposes \( P \) into a structured format amenable to translation.\\
 \( \mathcal{G} \): translates this structure into the SMT-LIB format.\\
 \( \mathcal{T} \): applies the theorem prover to find a solution that satisfies the logical constraints.\\
 \( \mathcal{E} \): evaluates this solution to determine its adequacy in solving the puzzle's clues.\\
 \( \mathcal{F} \): adjusts the translation based on the evaluation, aiming to correct any errors or optimize the solution.
%\end{itemize}

\paragraph{Convergence Criteria} The convergence of this iterative process is governed by the Solver LLM-agent's evaluation function \( \mathcal{E} \), which assesses both the correctness of the solution against the domain-specific requirements and the presence of any syntactic or semantic errors detected by \( \mathcal{T} \). We assume that \( \mathcal{E} \) is a black-box function defined by the instructions given to the LLM. The loop terminates when \( \mathcal{E} \) returns a value indicating that the solution \( S_k \) sufficiently meets all puzzle requirements and contains no detectable errors, or when a maximum retry limit has been reached.

\paragraph{Optimization and Refinement} Each iteration through the feedback loop serves to progressively refine the solution, optimizing the representation and alignment with the puzzle's constraints. This optimization process is critical for moving the solution towards a local optimum, where no further improvements can be detected by \( \mathcal{E} \) or \( \mathcal{F} \).

\section{Experimental Setup}

To comprehensively evaluate \tool's performance, we examine its effectiveness across 114 Zebra Puzzles. 
Our assessment emphasizes \tool's capability to solve the puzzles 
%with
using the different 
%KM with "different capabilities" this was too vague for me. I think what you are referring to here is an ablation study of the agents.
%capabilities. 
agents.

\subsection{Selection of Logic Puzzles}

We compiled two datasets to evaluate the problem-solving capabilities of our agent-centric approach. The first dataset, sourced from GitHub\footnote{\url{https://github.com/ross-nordstrom/LogicSolver/tree/master/data}}, contains 59 Zebra puzzles involving entity-attribute matching.
We further curated 55 additional puzzles from different sources from the Web and manually cross-checked them to determine they are valid zebra problems.
%prove validity emphasizing both the arrangement and entity assignments. 
%The second dataset~\rayb{where did you get it?} includes 50 puzzles 
% Problems were selected based on complexity levels marked as "easy" or "hard" across both datasets.
% \rayb{are we separating hard or easy puzzles in our eval?}

\subsection{Agent Configuration}

We experimented with three different LLMs: GPT-4, GPT-3.5, and Llama3-8b. We used z3 as the automated theorem solver. Our total cost across all experiments was approximately 2500 USD.

\paragraph{Number of Retries} In our experimental setup, we initially conduct the feedback loop once. To enhance performance and address syntactical errors in the final output, we implement an additional cold-start retry mechanism if we reach the action-limit without an error-free solution. This involves restarting the workflow from scratch with an increased temperature.

\paragraph{Response Limit} To limit the conversation length and prevent hallucination, we define a maximum number of actions that the LLM-agent can take. All experiments performed allow the LLM to perform up to 4 actions; this limit is reset if the puzzle-solving task is retried.

\subsubsection{Grading}

To assess solution accuracy, we created an autograder LLM-agent that provides a numeric grade to every solution generated by the solving agent. Each assignment is worth 1 point. In order to evaluate the reliability of this autograder, we also conducted a user study where a subset of the problems were regraded by humans and then compared to the autograder's results.

\paragraph{Autograding}
GPT-4o was used for autograding. It received the ground-truth answer, final SMT-LIB output, and conversation history to assess the consistency and correctness of each solution. For each problem, the model compared the logical assignments produced by the solving agent against the reference assignments, producing a final accuracy score.

To demonstrate the autograding process, 
%KM removing unnecessary words
%let us 
consider a scenario where the output of the SMT-LIB solver is evaluated against a pre-defined answer key. The solver's output and subsequent interpretation by the grader are detailed below.

\subparagraph{SMT-LIB Solver Output}
Below is a sample SMT solution for the logic grid puzzle given in Fig. \ref{fig:example}.
\begin{verbatim}
; 1 is Brazilian, 2 is German
; 3 is American
(define-fun H1_Color () String "Blue")
(define-fun H1_N () Int 1)
(define-fun H1_Anml () String "Cats")
(define-fun H1_Sp () String "Football")
(define-fun H2_Color () String "Green")
(define-fun H2_N () Int 3)
(define-fun H2_Anml () String "Dogs")
(define-fun H2_Sp () String "Basketball")
(define-fun H3_Color () String "Red")
(define-fun H3_N () Int 2)
(define-fun H3_Anml () String "Fishes")
(define-fun H3_Sp () String "Baseball")
\end{verbatim}

\subparagraph{Ground Truth Answer}
\begin{itemize}[itemsep=-1ex]
    \item House 1: Blue, Brazilian, Fishes, Football
    \item House 2: Green, American, Cats, Baseball
    \item House 3: Red, German, Dogs, Basketball
\end{itemize}

The autograder evaluates the solution by mapping the SMT-LIB output to the expected results either using contextual clues or an explicitly defined lookup table, which would be defined in the SMT-LIB comments, converting function definitions into comparable assignments, as in Table \ref{tab:validation_results}.

\begin{table}[ht]
\centering
\resizebox{0.9\columnwidth}{!}{
\begin{tabular}{lll}
\toprule
\textbf{Entity} & \textbf{Assignment} & \textbf{Result} \\
\midrule
House 1 & Color: Blue & \checkmark \\
House 1 & Nationality: Brazilian & \checkmark \\
House 1 & Animal: Cats & \xmark \\
House 1 & Sport: Football & \checkmark \\
\midrule
House 2 & Color: Green & \checkmark \\
House 2 & Nationality: American & \checkmark \\
House 2 & Animal: Dogs & \xmark \\
House 2 & Sport: Basketball & \xmark \\
\midrule
House 3 & Color: Red & \checkmark \\
House 3 & Nationality: German & \checkmark \\
House 3 & Animal: Fishes & \xmark \\
House 3 & Sport: Baseball & \xmark \\
\bottomrule
\end{tabular}
}
\caption{Validation Results}
\label{tab:validation_results}
\end{table}

\subparagraph{Partial Scoring (PS)}
Each correct match between the SMT-LIB output and the answer key earns a point. The autograder agent also calculates the total number of assignments which is equal to the number of points it is possible to receive. In this example, the colors and nationalities are correct, but the animals and two of the sports are wrong, thus:

\begin{equation*}
    \text{Partial Score} = \frac{\text{Correct Matches}}{\text{Total Matches}} = \frac{7}{12} = 0.58
\end{equation*}

If the animals and sports had been chosen correctly, the score would be 1.

\paragraph{Manual User Study Grading}
A separate user study manually graded 50 solutions from the state-of-the-art workflow, 35 solutions from a non-optimal variant, and 20 solutions from the naive approach. Though it was impractical to have all of the thousands of solutions that the LLM-agent generated be hand-graded, this user study allows us to quantify the correctness of our results and verify that our autograder correlates well with the ground-truth grades.

The manual grading team included five undergraduate computer science students and one master's student. We then used their manual grades to capture various statistical measures of similarity between human grading and LLM grading; these stats are explained in the "Results" section.

A large percentage of the attempted solutions included explicit lookup tables, making these solutions significantly more time-consuming to grade (see "SMT-LIB Solver Output" and "Answer Key" above). The lookup table could appear anywhere in the generated text, which comprises multiple blocks of SMT-LIB code, errors, and intermediate SMT models. We therefore do not include them in our user study.

\section{Results}
\label{sec:results_analysis}
This analysis is structured around four key research questions: Firstly, we examine the baseline performance of different LLMs in solving logic puzzles without solver assistance to understand their intrinsic problem-solving capabilities. Secondly, we assess the improvements in accuracy and problem-solving completeness when integrating solver feedback, evaluating how external theorem provers enhance LLM effectiveness. Thirdly, we explore the impact of using a decomposition agent, analyzing whether segmenting puzzles into simpler components before solving improves overall solution quality. Fourth, we conduct a user study to evaluate our LLM-Grader and substantiate the validity of our results.

\subsection{\tool Performance over Baselines}
To establish a baseline, we first evaluate the performance of LLMs without the assistance of a solver by asking the LLM to solve the logic grid puzzle.
This baseline configuration yields mediocre puzzle-solving accuracy, as detailed in Table~\ref{tab:baseline_performance1}. We report both the average partial score, given by the "Avg. PS" column, and the number of puzzles solved fully correctly, given by the "\#Solved" column. For instance, GPT-4 under a baseline achieves an average partial score of 52.4\% and solves 27/114 logic grid puzzles completely correctly.

The effectiveness of the LLM-agent workflow increases markedly when solver feedback is incorporated.  As shown in Table~\ref{tab:solver_integration}, the integration of theorem prover feedback, without retries and under a deterministic generation setting (temperature = 0), increases GPT-4's average partial score to 0.687 from baseline of 0.524 ($\Delta=31.1$\%).
The inclusion of a decomposition agent further improves this to 0.700 ($\Delta=33.6$\%). 
In terms of the total number solutions that can be completely solved, GPT4 with solver solves up to 133.33\% more problems than the baseline settings. GPT-3.5 shows a similar positive trend. 
%improvement, solving 29\% more problems completely correctly when the SMT feedback is introduced and 41\% more problems completely correctly when a decomposition agent is added. 

Llama3's improvement is more subtle; we believe this is because its fewer number of parameters limits its ability to generate syntactically correct SMT-LIB code. This theory is supported by the fact that in every Llama3 experiment, no less than 50 final solutions contained errors, whereas in every GPT-4 or GPT-3.5 experiment, the number was no more than 42. Nonetheless, Llama3 can also improve the total number of correct solutions by 50\% over baseline. 

\subsection{\tool Performance under Different Settings}
For the variable temperature experiments, we set the model temperature to zero and increased it if the solution contained errors.  While this approach provides the flexibility to bypass a solution if the deterministic solution is erroneous, it risks generating less stable solutions that may inadvertently replace syntactically incorrect yet valid solutions with syntactically correct but logically flawed ones. This phenomenon is particularly pronounced in models with fewer parameters, where the performance tends to decline with the introduction of retries. For example, under variable temperature conditions with retries, GPT-4 maintains a high accuracy rate of 76.1\%, while Llama3's accuracy degrades to 43.6\%.

The addition of a decomposition agent to the SMT-integrated LLM-agent yielded mixed results. For both GPT-4 and GPT-3.5, the average partial score slightly improved, in both cases by less than 5.5\%, while the number solved fully correctly rose modestly (by 4 and 2 problems, respectively). However, Llama-3's average partial score declined by less than 5\% and it was able to solve 3 fewer problems than with just SMT integration. Because of the relatively small differences in all cases, more experimentation is needed to determine when decomposition increases performance.

% Tables for experimental setups
\begin{table}[htbp]
\centering
%\caption{Baseline Performance of LLMs Without Solver Integration. All experiments were run with a temperature of 0. }
\begin{tabular}{|l|c|c|c|c|}
\toprule
\textbf{Model}  & \textbf{T} & \textbf{D} & \textbf{Avg. P.S} & \textbf{\#Solved} \\
\midrule
Llama3-8b &  0 & \xmark & 0.47 & 14 (12.3\%) \\
GPT-3.5 &   0 & \xmark & 0.471 & 17 (15.0\%)\\
GPT-4 &   0 & \xmark & 0.524 & 27 (23.7\%) \\
\bottomrule
\end{tabular}
\caption{Baseline Performance of LLMs Without Solver Integration.The "D" column indicates if a decomposition agent was present in the workflow. The "T" column indicates Temperature. }
\label{tab:baseline_performance1}
\end{table}

\begin{table}[htbp]
\centering
\resizebox{\columnwidth}{!}{
\begin{tabular}{|l|r|r|r|r|r|}
\toprule
\textbf{Model} & \textbf{T} & \textbf{D} & \textbf{Avg. PS} & \textbf{\#Solved} & \textbf{$\Delta\#$Solved} \\
\midrule
Llama3-8b & 0 & \xmark & 0.496 & 21 (18.4\%)  & 50.0\% \\
GPT-3.5 & 0 & \xmark & 0.493 & 22 (19.3\%)  & 29.4\% \\
GPT-4 & 0 & \xmark & 0.687 & 59 (51.8\%) & 118.5\% \\
\midrule
Llama3-8b & Var. & \xmark & 0.436 & 15 (13.2\%) &  7.1\% \\
GPT-3.5 & Var. & \xmark & 0.484 & 24 (21.0\%) &  41.2\% \\
GPT-4 & Var. & \xmark & 0.761 & 72 (63.2\%) & 166.7\% \\
\midrule
Llama3-8b & 0 & \checkmark & 0.468 & 18 (15.8\%)  & 28.6\% \\
GPT-3.5 & 0 & \checkmark & 0.520 & 24 (21.0\%) &  41.2\% \\
GPT-4    & 0 &  \checkmark & 0.700 & 63 (55.3\%) &  133.3\% \\
\bottomrule
\end{tabular}
}
\caption{Enhancements from Solver Integration with Percentage Improvement Over Baseline. The "D" column indicates if a decomposition agent was present in the workflow. The "T" column indicates Temperature.}
\label{tab:solver_integration}
\end{table}

\subsection{Manual Analysis of Grading}

Based on our user study, our LLM-based grading systems demonstrate high accuracy across a variety of models and settings. The system maintains consistent scoring accuracy, with exact match rates exceeding 78\% across all tested scenarios. 

To evaluate the LLM-grader, we employed various statistical measures: (i) {Avg.~Abs.~Diff.~}: The average magnitude of the difference between the partial score given by the LLM-grader and the human evaluator. (ii) {Avg.~Rel.~Diff.~}: The expected percent difference between the partial score given by the LLM-grader and the human evaluator.  \% problems for which the LLM-grader (iii) overestimated and (iv) underestimated the partial score provided by the human evaluator. (v) \% problems for which the LLM-grader and the human evaluator gave exactly the same partial score, and (vi) Joint Full Credit: The count of problems for which both the user and the LLM assigned full credit, normalized over the total number of problems that either party marked for full credit. This metric helps in understanding the extent of agreement in the grading of solutions between the human and machine evaluators.

The metric of "Joint Full Credit," which consistently registers above 85\%, serves as a robust indicator of  the LLM-grader's capability to accurately assess fully correct solutions as demonstrated by the level of agreement between the LLM and a human grader. Additionally, the analysis indicates a propensity for the grader to overestimate the score of the LLM without SMT integration, whereas the integration of SMT tends to result in slight underestimations by the grader. This observation suggests that the integration of SMT and a feedback based loop may contribute more significantly to performance improvements than the raw grading differentials indicate.

\begin{table}[ht]
\centering
\label{tab:merged_user_study}
\resizebox{\columnwidth}{!}{
\begin{tabular}{lrrr}
\toprule
                   & \textbf{GPT-4} & \textbf{GPT-3.5} & \textbf{GPT-4}  \\
                   & \textbf{SMT+D} & \textbf{Naive} & \textbf{SMT} \\
\textbf{Statistic} & \textbf{(50)} & \textbf{(20)} & \textbf{ (35)} \\
\midrule
Exact Match (\%) & 78.26 & 78.94 & 82.35 \\
Avg. Abs.~ Diff & 0.056 & 0.040 & 0.117 \\
Avg. Rel. Diff (\%) & -3.55 & +13.80 & +2.92 \\
LLM Overestimated (\%) & 13.04 & 21.05 & 11.76 \\
LLM Underestimated (\%) & 8.70 & 0.00 & 5.88 \\
Joint Full Credit (\%) & 89.19 & 100 & 86.96 \\
Spearman Correlation  & 0.73 & 0.95 & 0.70 \\
\bottomrule
\end{tabular}
}
\caption{User Study Statistics comparing different experimental setups. The first column is GPT-4 with SMT integration and the decomposition (D) agent over 50 problems. The second column is GPT-3.5 without SMT integration over 20 problems. The third column is GPT-4 with just SMT integration over 35 problems. All problems were graded by both the manual grader and the LLM.}
\end{table}

\section{Background and Related Work}

The idea of using artificial intelligence with sensing or feedback components to detect and self-correct errors has been well-studied. Bowen and Kang \cite{bowen1988using} used memory of previous statements to correct and formalize fuzzy natural language sentences. Self-driving cars use sensory technology to correct errors in speed and output current \cite{chen2020survey}. More recently, large language models (LLMs) have shown they are capable of learning from real-time user feedback \cite{hancock2019feedback, izadi2024error} outside of training.

The concept of agent-centric LLM agents, as discussed in recent literature revolves around creating systems (usually backed by LLMs) that can act independently in diverse environments, both physical and virtual \citep{wang2024survey}. This framework shifts the focus from passive systems to proactive entities capable of dynamic interaction and problem-solving. In these models, agents are designed to perceive and react to multi-modal data, integrating visual, auditory, and textual input to generate appropriate actions in real time. The most tangible benefit of this framework is feedback, which can take the form of a physical environment, error correction, or manual input \citep{durante2024}.

Significant work has been done to apply this framework to solving text-based puzzles.  \citet{zhou2023} used a process called Language Agent Tree Search (LATS), which integrates planning, reasoning, and acting within LLMs to decompose and solve a high-level reasoning task. ~\citet{gao2023pal} showed that generating intermediate representations as Python programs allowed small LLMs to outperform much larger ones. Logic-LM and SatLM both used LLMs to generate formal representations of general natural language problems and used off the shelf theorem provers to generate answers \citep{pan2023, ye2023}. While none of these approaches focus on Zebra puzzles, they each show that LLMs perform better when used as agents in a formally grounded system.%~\rayb{Beyond solving Zebra not puzzle, how are they different?}.

%\kmnote{This paragraph is nice but longer than one would usually find in related work. Keep in mind as a potential for cuts if we need more space.}
Research has shown that natural language cannot be mapped one-to-one with a formal space due to inherent ambiguities \citep{osama2023}. For our approach, it was thus vital to create an agent that can take into account context and background knowledge to figure out the correct translation into a formal space.
Even if the clues were perfectly translated as they are presented, a formal solver will not be able to generate a fully correct solution without additional encoding by the problem translator of this general context. Our approach is different from prior agent approaches in that we use a structured symbolic space (SMT-LIB) but use the syntactic and semantic feedback from an automated theorem prover for analysis in an LLM agent. We also provide a conceptual framework to understand LLM interaction with the automated theorem prover and its generated text as an agent.

%Research indicates that natural language cannot be perfectly mapped to a formal space due to inherent ambiguities \citep{osama2023}. Thus, our approach creates an agent that incorporates context and background knowledge for accurate translation into a formal space. Unlike prior methods, we use a structured symbolic space (SMT-LIB) and integrate feedback from an automated theorem prover into an LLM agent. Additionally, we offer a framework for understanding the interaction between the LLM and the theorem prover.

\section{Conclusion}
\label{sec:conclusion}
%This research shows the effectiveness of a multi-agent framework that bolsters the performance of large language models (LLMs) in solving logic puzzles and other natural language task. The inter-agent critique mechanism plays a crucial role. Through dialogues, agents can critique and refine each other's contributions, which leads to more accurate and consistent results. This process of mutual review and adjustment cultivates a collaborative intelligence that significantly enhances the quality of solutions. The development of an autograder with a verified correlation to human evaluation is also an important benchmark to this task.

This research shows the effectiveness of a multi-agent LLM and SMT framework that bolsters the performance of large language models (LLMs) in solving logic puzzles and other natural language tasks. Our work demonstrates the importance of integrating LLMs and SMTs in the task, boosting performance over an LLM alone. We also show that the inter-agent critique mechanism plays a crucial role. 
Through dialogues, agents  critique and refine each other's contributions, which leads to more accurate and consistent results. 
%This process of mutual review and adjustment cultivates a collaborative intelligence that significantly enhances the quality of solutions. 
The development of an autograder, with a verified correlation to human evaluation, played a role in the feedback mechanism by indicating when a solution was not judged logically correct and also enabled iterative development of our approach. Our findings suggest that structured planning and agent-feedback greatly enhance LLMs’ capability to solve logical problems.

%is also an important benchmark to this task.

% Looking ahead, there are several avenues for further research. These include optimizing retry mechanisms to discover more effective solution pathways, which could be informed by approaches such as Program-of-Thoughts and Graph-of-Thoughts-Rationale \citep{chen2023program},~\citep{Besta_2024}. Moreover, increasing the agent-environment size and the length of the feedback loop would likely give the solving agent a higher chance of correcting itself; doing this would requiring increasing the actual context limit and the effective context limit where the LLM-agent can remember past strategies properly.

 Looking ahead, further research could optimize retry mechanisms for discovering more effective solutions, informed by approaches like Program-of-Thoughts and Graph-of-Thoughts-Rationale \citep{chen2023program,Besta_2024}. Additionally, increasing the agent-environment size and the feedback loop length would enhance the solving agent's self-correction capabilities by expanding the actual and effective context limits for remembering past strategies.

\section{Limitations}
\label{sec:limitations}

This study, while advancing our understanding of LLMs in solving logic puzzles, has several limitations that warrant further investigation. Firstly, the experiments were confined to only three models: GPT-4, GPT-3.5, and Llama3-8b. 
%KM This sounds too negative and reviewers pick up on negative comments
%This restriction raises the question of 
Investigation of 
generalizability across different LLMs is warranted, especially because our performance gains occurred mainly in the GPT family.

Secondly, our approach relied on specific prompt constructions for both the grader and the solver agents. There exists a possibility that alternative prompting strategies could yield more accurate or efficient problem-solving and grading results. Further research is needed to explore and optimize these prompts to fully leverage the potential of LLMs in this domain.

Additionally, our user study  inherently carries some uncertainty regarding its correlation to actual problem-solving performance. Solutions involving complex lookup tables were excluded from the user study due to how time consuming they were to grade, which might affect the study's comprehensiveness and the general applicability of our findings.

Lastly, our bank of logic grid puzzles used in this study was somewhat limited in both size-- we used 114 problems-- and range of difficulty. The majority of puzzles used were subjectively rated as medium difficulty. Extending this research to include a larger and more varied dataset would verify the usefulness of our findings.

\section{Ethics Statement} \label{sec:ethics-statement}

\noindent \textbf{Use of Generative AI.} Generative models carry ethical risks, including the potential to produce harmful content or content that closely mirrors pre-training data. However, we are using the generative models to solve puzzles rather than showing their direct output. Thus, such risk is minimal. 

\noindent \textbf{Compute.} Employing deep learning models is computationally intensive and can have environmental implications. However, as no models were trained as part of this research, the computational impact remains relatively low. 

\noindent \textbf{Human Evaluator.} We use only 6 human evaluators who are undergraduate/masters students in the lab environment and were given full disclosure about the nature of the study and its unpaid nature. No ethical violations were committed in such setting.

\section*{Acknowledgments}

We will add acknowledgement in the non-anonymous version of the paper. 

\bibliography{latex/custom}

\appendix

\section{Appendix}
\subsection{Example Puzzle Workflow}
\label{sec:appendix}
We begin by inputting the puzzle and its entities into the Solver LLM-Agent (in this case, GPT-4).

\begin{Verbatim}[fontsize=\small]
first, second, third, fourth
Bridget, Kermit, Ophelia, Stretch
#105, #118, #126, #128

The ostrich that finished second was #128.
First place was either #118 or #126.
Third place was either #126 or Bridget.
Ophelia finished second.
Stretch finished 2 places after Kermit.
\end{Verbatim}

The solver agent responds by attempting to translate the puzzle and its clues into SMT-LIB:

\begin{Verbatim}[fontsize=\small]
(set-logic QF_LIA)

; Place of each ostrich
(declare-const Bridget_Place Int)
(declare-const Kermit_Place Int)
(declare-const Ophelia_Place Int)
(declare-const Stretch_Place Int)

; Number of each ostrich
(declare-const Bridget_Number Int)
(declare-const Kermit_Number Int)
(declare-const Ophelia_Number Int)
(declare-const Stretch_Number Int)

(assert (and
(or (= Bridget_Place 1)
(= Bridget_Place 2)
(= Bridget_Place 3) (= Bridget_Place 4))
(or (= Kermit_Place 1)
(= Kermit_Place 2)
(= Kermit_Place 3) (= Kermit_Place 4))
(or (= Ophelia_Place 1)
(= Ophelia_Place 2)
(= Ophelia_Place 3) (= Ophelia_Place 4))
(or (= Stretch_Place 1) 
(= Stretch_Place 2)
(= Stretch_Place 3) (= Stretch_Place 4))
))

(assert (and
(or (= Bridget_Number 105)
(= Bridget_Number 118)
(= Bridget_Number 126) 
(= Bridget_Number 128))
(or (= Kermit_Number 105) 
(= Kermit_Number 118)
(= Kermit_Number 126) 
(= Kermit_Number 128))
(or (= Ophelia_Number 105)
(= Ophelia_Number 118) 
(= Ophelia_Number 126) 
(= Ophelia_Number 128))
(or (= Stretch_Number 105)
(= Stretch_Number 118) 
(= Stretch_Number 126)
(= Stretch_Number 128))
))

; The ostrich that finished second was #128.
(assert (or (and (= Ophelia_Place 2) 
(= Ophelia_Number 128))
            (and (= Bridget_Place 2)
            (= Bridget_Number 128))
            (and (= Kermit_Place 2)
            (= Kermit_Number 128))
            (and (= Stretch_Place 2)
            (= Stretch_Number 128))))

; The ostrich that finished first was either #118 or #126.
(assert (or (and (= Ophelia_Place 1)
(or (= Ophelia_Number 118)
(= Ophelia_Number 126)))
            (and (= Bridget_Place 1)
            (or (= Bridget_Number 118)
            (= Bridget_Number 126)))
            (and (= Kermit_Place 1)
            (or (= Kermit_Number 118)
            (= Kermit_Number 126)))
            (and (= Stretch_Place 1)
            (or (= Stretch_Number 118)
            (= Stretch_Number 126)))))

; The ostrich that finished third was
; either #126 or Bridget.
(assert (or (and (= Ophelia_Place 3)
(= Ophelia_Number 126))
            (and (= Bridget_Place 3)
            (or (= Bridget_Number 126)
            (= Bridget_Number 105)
            (= Bridget_Number 118) 
            (= Bridget_Number 128)))
            (and (= Kermit_Place 3)
            (= Kermit_Number 126))
            (and (= Stretch_Place 3)
            (= Stretch_Number 126))))

; Ophelia finished second.
(assert (= Ophelia_Place 2))

; Stretch finished 2 places after Kermit.
(assert (= Stretch_Place (+ Kermit_Place 2)))

; Distinct places
(assert (distinct Bridget_Place 
Kermit_Place Ophelia_Place Stretch_Place))

; Distinct numbers
(assert (distinct Bridget_Number
Kermit_Number Ophelia_Number Stretch_Number))
\end{Verbatim}

This SMT-LIB code is then fed into z3, which produces the following output:
\begin{Verbatim} [fontsize=\small]
(error "line 15
column 0: invalid command, '(' expected")
(error "line 15
column 2: unexpected character")
(error "line 15
column 77: unexpected character")
(error "line 15
column 88: unexpected character")
(error "line 15
column 117: unexpected character")
(error "line 15
column 136: unexpected character")
(error "line 15
column 142: unexpected character")
(error "line 31
column 0: invalid command, '(' expected")
(error "line 31
column 7: unexpected character")
(error "line 57
column 0: invalid command, '(' expected")
(error "line 57
column 7: unexpected character")
(error "line 65
column 0: invalid command, '(' expected")
sat
(
  (define-fun Ophelia_Number () Int
    128)
  (define-fun Stretch_Place () Int
    3)
  (define-fun Stretch_Number () Int
    126)
  (define-fun Bridget_Place () Int
    4)
  (define-fun Kermit_Number () Int
    118)
  (define-fun Bridget_Number () Int
    105)
  (define-fun Ophelia_Place () Int
    2)
  (define-fun Kermit_Place () Int
    1)
)
\end{Verbatim}

This output is fed back into the Solver LLM-Agent, which then produces the following modified SMT-LIB code:

\begin{Verbatim}[fontsize=\small]
(set-logic QF_LIA)

; Place of each ostrich
(declare-const Bridget_Place Int)
(declare-const Kermit_Place Int)
(declare-const Ophelia_Place Int)
(declare-const Stretch_Place Int)

; Number of each ostrich
(declare-const Bridget_Number Int)
(declare-const Kermit_Number Int)
(declare-const Ophelia_Number Int)
(declare-const Stretch_Number Int)

; The places are in the set {1, 2, 3, 4}
(assert (and
(or (= Bridget_Place 1) 
(= Bridget_Place 2) (= Bridget_Place 3) 
(= Bridget_Place 4))
(or (= Kermit_Place 1) 
(= Kermit_Place 2) (= Kermit_Place 3) 
(= Kermit_Place 4))
(or (= Ophelia_Place 1) 
(= Ophelia_Place 2) (= Ophelia_Place 3) 
(= Ophelia_Place 4))
(or (= Stretch_Place 1) 
(= Stretch_Place 2) (= Stretch_Place 3) 
(= Stretch_Place 4))))

; Numbers are in the set {105, 118, 126, 128}
(assert (and
(or (= Bridget_Number 105) 
(= Bridget_Number 118) (= Bridget_Number 126) 
(= Bridget_Number 128))
(or (= Kermit_Number 105) 
(= Kermit_Number 118) (= Kermit_Number 126) 
(= Kermit_Number 128))
(or (= Ophelia_Number 105) 
(= Ophelia_Number 118) (= Ophelia_Number 126) 
(= Ophelia_Number 128))
(or (= Stretch_Number 105) 
(= Stretch_Number 118) (= Stretch_Number 126) 
(= Stretch_Number 128))))

; The ostrich that finished second was #128.
(assert (or (and (= Ophelia_Place 2) 
(= Ophelia_Number 128))
            (and (= Bridget_Place 2) 
            (= Bridget_Number 128))
            (and (= Kermit_Place 2) 
            (= Kermit_Number 128))
            (and (= Stretch_Place 2) 
            (= Stretch_Number 128))))

; The ostrich that finished first was 
; either #118 or #126.
(assert (or (and (= Ophelia_Place 1)
(or (= Ophelia_Number 118) 
(= Ophelia_Number 126)))
            (and (= Bridget_Place 1)
            (or (= Bridget_Number 118) 
            (= Bridget_Number 126)))
            (and (= Kermit_Place 1)
            (or (= Kermit_Number 118) 
            (= Kermit_Number 126)))
            (and (= Stretch_Place 1) 
            (or (= Stretch_Number 118)
            (= Stretch_Number 126)))))

; The ostrich that finished third
; was either #126 or Bridget.
(assert (or (and (= Ophelia_Place 3)
(= Ophelia_Number 126))
            (and (= Bridget_Place 3) 
            (or (= Bridget_Number 126) 
            (= Bridget_Number 105) 
            (= Bridget_Number 118) 
            (= Bridget_Number 128)))
            (and (= Kermit_Place 3) 
            (= Kermit_Number 126))
            (and (= Stretch_Place 3)
            (= Stretch_Number 126))))

; Ophelia finished second.
(assert (= Ophelia_Place 2))

; Stretch finished 2 places after Kermit.
(assert (= Stretch_Place (+ Kermit_Place 2)))

; Distinct places
(assert (distinct 
Bridget_Place Kermit_Place 
Ophelia_Place Stretch_Place))

; Distinct numbers
(assert (distinct 
Bridget_Number Kermit_Number
Ophelia_Number Stretch_Number))

(check-sat)
(get-model)
\end{Verbatim}

Z3 produces the following output, which is then determined to be both error-free and consistent:

\begin{verbatim}

sat
(
  (define-fun Ophelia_Number () Int
    128)
  (define-fun Stretch_Place () Int
    3)
  (define-fun Stretch_Number () Int
    126)
  (define-fun Bridget_Place () Int
    4)
  (define-fun Kermit_Number () Int
    118)
  (define-fun Bridget_Number () Int
    105)
  (define-fun Ophelia_Place () Int
    2)
  (define-fun Kermit_Place () Int
    1)
)
\end{verbatim}

\subparagraph{Ground Truth Answer}
\begin{itemize}[itemsep=-1ex]
    \item Kermit: First, \#118
    \item Ophelia: Second, \#128
    \item Stretch: Third, \#126
    \item Bridget: Fourth, \#105
\end{itemize}

The autograder evaluates the solution by mapping the SMT-LIB output to the expected results either using contextual clues or an explicitly defined lookup table, which would be defined in the SMT-LIB comments, converting function definitions into comparable assignments, as in Table \ref{tab:validation_results2}. In this case, the solution gets full credit.

\begin{table}[ht]
\centering
\caption{Validation Results}
\label{tab:validation_results2}
\begin{tabular}{ll}
\toprule
\textbf{Entity} & \textbf{Result} \\
\midrule
Kermit & Place: First (Correct) \\
Kermit & Number: 118 (Correct) \\
Ophelia & Place: Second (Correct) \\
Ophelia & Number: 128 (Correct) \\
Stretch & Place: Third (Correct) \\
Stretch & Number: 126 (Correct) \\
Bridget & Place: Fourth (Correct) \\
Bridget & Number: 105 (Correct) \\
\bottomrule
\end{tabular}
\end{table}

\subsection{User Study Instructions}
The following instructions were presented to our manual graders before they began grading. The full UI can be found at~\url{https://github.com/shmublu/anon_emnlp/tree/main} by running the "autograder\_flask.py" file.

\includegraphics[width=0.5\textwidth]{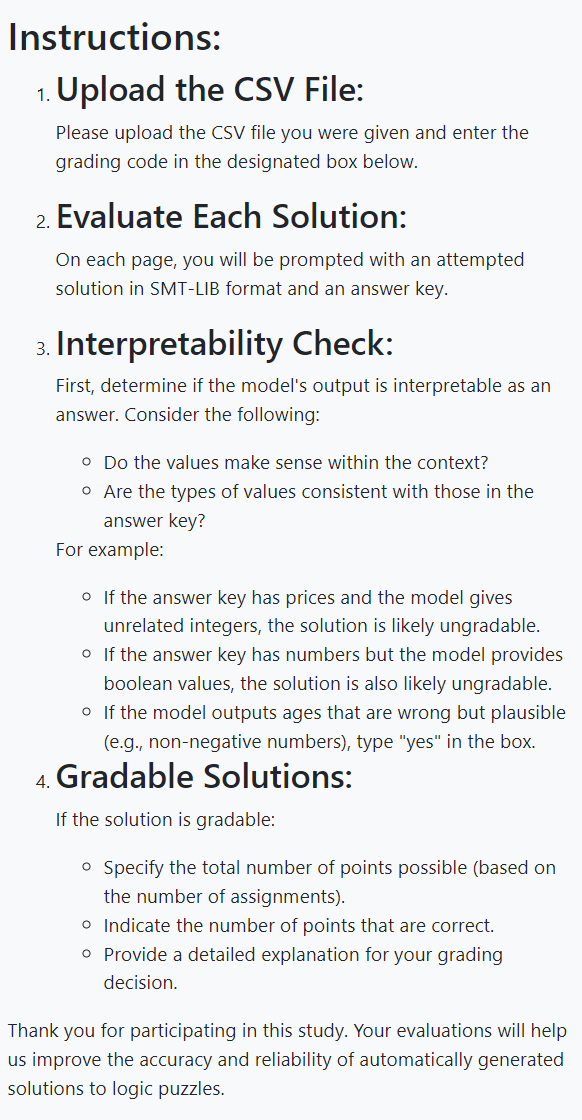}

\end{document}